\documentclass[twocolumn,amsmath, pra, showpacs]{revtex4}%
\usepackage{amsfonts}
\usepackage{amsmath}
\usepackage{amssymb}
\usepackage{graphicx}%
\providecommand{\U}[1]{\protect\rule{.1in}{.1in}}

\begin{document}
\title{Generation of arbitrary symmetric entangled states with conditional linear
optical coupling}
\author{A. V. Sharypov}
\affiliation{Kirensky Institute of Physics, 50 Akademgorodok, Krasnoyarsk, 660036, Russia}
\affiliation{Siberian Federal University, 79 Svobodny Ave., Krasnoyarsk, 660041, Russia}
\author{Bing He}
\affiliation{Institute for Quantum Information Science, University of Calgary, Calgary,
Alberta T2N 1N4, Canada}


\begin{abstract}
An approach for generating the entangled photonic states $\left\vert \Psi
_{1},\Psi_{2}\right\rangle \pm\left\vert \Psi_{2},\Psi_{1}\right\rangle $ from
two arbitrary states $|\Psi_{1}\rangle$ and $|\Psi_{2}\rangle$ is proposed.
The protocol is implemented by the conditionally induced beam-splitter
coupling which leads to the selective swapping between two photonic modes.
Such coupling occurs in a quantum system prepared in the superposition of two
ground states with only one of them being involved in the swapping. All the
entangled states in the category, such as entangled pairs of coherent states
or Fock states (N00N states), can be efficiently produced in the same way by
this method.

\end{abstract}
\maketitle

\section{introduction}

Bipartite symmetric entangled states refer to a generic type in the form
$\left\vert \Psi_{1},\Psi_{2}\right\rangle \pm\left\vert \Psi_{2},\Psi
_{1}\right\rangle $ up to a normalization factor. Such entangled states
include the symmetric entangled coherent states (SECSs) $\left\vert
\alpha,\beta\right\rangle \pm\left\vert \beta,\alpha\right\rangle $
\cite{SandersRev} and the N00N states $\left\vert N,0\right\rangle
\pm\left\vert 0,N\right\rangle $ \cite{noon1, noon2}. Both of them have found
important applications in quantum metrology; see, e.g. \cite{metro, metro2}. A
SECS of light fields can be transformed to a photonic Schr\"{o}dinger cat
state $|\gamma\rangle\pm|-\gamma\rangle$ \cite{cat} simply by a beam-splitter
(BS) operation. Cat states of matter wave and even light field have been
experimentally demonstrated \cite{cat1, cat2, cat3}, but a photonic one with
the sufficiently large size $|\gamma|$ is still beyond the reach.

Since the seminal work of Yurke and Stoler \cite{y-s}, the application of Kerr
nonlinearity has been suggested as the direct way to entangle light fields or
construct photonic cat states \cite{SandersRev}. Realizing strong coupling
between photons via the suitable nonlinear media is, however, a rather
difficult task. This barrier stimulates the parallel researches on creating
the approximate states by squeezing (see, e.g. \cite{squeeze, squeeze2} and
the reference of \cite{cat}) and exploring the proper use of weak Kerr
nonlinearity (see, e.g. \cite{jeong, k-p, h-09}).

A less noticed problem with Kerr nonlinearity and squeezing is the
availability of their single-mode versions, which are the basis for all
relevant schemes thus far. A realistic photonic pulse carries multiple modes
represented by the field operator $\hat{\mathcal{E}}(z)=\sum_{k}\hat{a}%
_{k}e^{ikz}$ (in one-dimensional space for illustration). For instance, under
the action of a multi-mode self-Kerr Hamiltonian $\int dz\big(\hat
{\mathcal{E}}^{\dagger}(z)\hat{\mathcal{E}}(z)\big)^{2}$ of the unit coupling
constant or its equivalent form $\sum_{k1,k2,k3}\hat{a}_{k1-k2+k3}^{\dagger
}\hat{a}_{k1}\hat{a}_{k2}^{\dagger}\hat{a}_{k3}$ in the wave-vector space, the
output states can be significantly different from the proper ones that should
have evolved under the sum of single-mode actions $\sum_{k}(\hat{a}%
_{k}^{\dagger}\widehat{a}_{k})^{2}$, even if the inputs are exactly
single-mode ones. This effect of mode entanglement or mode mixing has been
detailedly studied in \cite{g, h-12}. A consequence of the effect is a
vanishing or a very limited clean cross phase (similar to that obtained from
the single-mode cross-Kerr model) under highly demanding conditions
\cite{h-11-2,h-11}. On the other hand, a multi-mode squeezing action $\sum
_{k}(\hat{a}_{k}^{\dagger}\hat{a}_{-k}^{\dagger}+\hat{a}_{k}\hat{a}_{-k})$ of
one field as well deviates from its single-mode version. In contrast, the
multi-mode BS Hamiltonian $H_{BS}$ for two fields $\hat{\mathcal{E}}_{1}(z)$
and $\hat{\mathcal{E}}_{2}(z)$ takes the form $\int dz\{\hat{\mathcal{E}}%
_{1}^{\dagger}\hat{\mathcal{E}}_{2}(z)+\hat{\mathcal{E}}_{1}\hat{\mathcal{E}%
}_{2}^{\dagger}(z)\}=\sum_{k}(\hat{a}_{1,k}^{\dagger}\hat{a}_{2,k}+\hat
{a}_{1,k}\hat{a}_{2,k}^{\dagger})$, a sum of the individual mode actions. This
BS coupling enables a multi-mode photonic state to be transformed ideally like
a single-mode one, because the decomposable evolution operator $\exp
\{-itH_{BS}\}$ with respect to the wave-vector modes $k$ acts independently on
each mode.

In this paper we provide a method for generating arbitrary symmetric entangled
states out of light fields based only on such \textit{clean} BS coupling .
Unlike a common linear optical setup, the BS coupling we need acts
conditionally on the part of a superposition of quantum states at the same
spatial location. Below we will show how to produce a symmetric entangled
state with a conditional BS coupling and will give an example of the
realization of the given type of interaction in a proper quantum system.

\section{Protocol to entangle arbitrary input states}

From now on, we use the term \textit{mode} in the meaning of a single
wave-vector or a single frequency mode, since we will consider BS type
coupling only. The two arbitrary states $\left\vert \Psi_{1}\right\rangle $,
$\left\vert \Psi_{2}\right\rangle $ we will entangle are treated as the
single-frequency ones.

To entangle the two states, we also need an ancilla quantum system with two
stable states $\left\vert m\right\rangle $ and $\left\vert g\right\rangle $.
This system can be an atom, as well as an ion, a quantum dot or a
superconducting qubit. The ancilla system is initially in the state
$\left\vert m\right\rangle $, setting the initial state for the total system
as $\left\vert \Phi_{0}\right\rangle =\left\vert \Psi_{1},\Psi_{2}%
\right\rangle \left\vert m\right\rangle $. Then we perform a $\sigma^{x}%
$\ rotation between the $\left\vert m\right\rangle $ and $\left\vert
g\right\rangle $ and transfer the system to the superposition%
\begin{equation}
\left\vert \Phi_{1}\right\rangle =\left\vert \Psi_{1},\Psi_{2}\right\rangle
\left(  \left\vert m\right\rangle -i\left\vert g\right\rangle \right)
/\sqrt{2}.\label{State1}%
\end{equation}
Such $\sigma^{x}$ rotation can be realized by applying a resonant $\pi/2$
pulse to the transition $\left\vert m\right\rangle \rightarrow\left\vert
g\right\rangle $.%
\begin{figure}
[ptb]
\begin{center}
\includegraphics[
height=4.0222in,
width=3.0467in
]%
{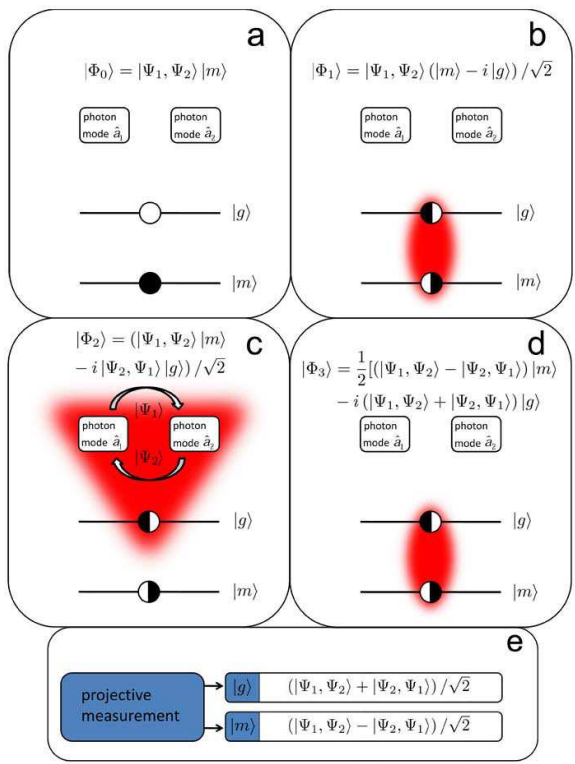}%
\caption{(color online). Conversion of the separable state into the two-mode
symmetrical entangled state. a) Initial state of the system. b) Preparation of
the ancilla system in the superposition state due to $\sigma^{x}$ rotation. c)
Creation of the superposition of the two events - field's state swapped and
does not swapped. d) Deentanglement from the ancilla system degree of freedom
due to $\sigma^{x}$ rotation. e) Projective measurement in the ancilla system
subspace collapse the photon wave function into the one of the two entangled
states.}%
\end{center}
\end{figure}
The above superposition of $\left\vert m\right\rangle $ and $\left\vert
g\right\rangle $ works as a logic control on the swapping between two input
photonic modes: the swapping between the photon modes is activated in the
$\left\vert g\right\rangle $ subspace and does not happen in the $\left\vert
m\right\rangle $ subspace. Such conditional swapping can be realized by the BS
transformation%
\begin{equation}
\left(
\begin{array}
[c]{c}%
\hat{a}_{1}(t)\\
\hat{a}_{2}(t)
\end{array}
\right)  =\left(
\begin{array}
[c]{cc}%
\cos\chi_{0}t & -ie^{i\phi}\sin\chi_{0}t\\
-ie^{-i\phi}\sin\chi_{0}t & \cos\chi_{0}t
\end{array}
\right)  \left(
\begin{array}
[c]{c}%
\hat{a}_{1}(0)\\
\hat{a}_{2}(0)
\end{array}
\right)  \label{idealBMHeis}%
\end{equation}
for the time $t_{s}=\pi/(2\chi_{0})$, where $\hat{a}_{i}$ is the annihilation
operator for the $i$-th mode and $\chi_{0}$ is the effective BS coupling
constant. The BS transformation can be implemented via the dispersive
parametric three-wave mixing (TWM) \cite{Serra, lin-08} or four-wave mixing
(FWM) \cite{Yavuz,SharypovFWMSuperconductors} process. The use of the
dispersive type of the interaction allows to avoid the decoherence of the
generated state due to incoherent scattering as during the BS interaction the
ancilla system is always preserved in its ground state and only the photon
states are changed. The conditional swapping results in the state
\begin{equation}
\left\vert \Phi_{2}\right\rangle =\left(  \left\vert \Psi_{1},\Psi
_{2}\right\rangle \left\vert m\right\rangle -i\left\vert \Psi_{2},\Psi
_{1}\right\rangle \left\vert g\right\rangle \right)  /\sqrt{2}.\label{State2}%
\end{equation}

Then, again we perform a $\sigma^{x}$\ rotation between $\left\vert
m\right\rangle $ and $\left\vert g\right\rangle $ to have them transformed as
$\left\vert m\left(  g\right)  \right\rangle \rightarrow\left\vert m\left(
g\right)  \right\rangle -i\left\vert g\left(  m\right)  \right\rangle $,
leading to the following state%
\begin{align}
\left\vert \Phi_{3}\right\rangle  &  =\frac{1}{2}[\left(  \left\vert \Psi
_{1},\Psi_{2}\right\rangle -\left\vert \Psi_{2},\Psi_{1}\right\rangle \right)
\left\vert m\right\rangle \nonumber\\
&  -i\left(  \left\vert \Psi_{1},\Psi_{2}\right\rangle +\left\vert \Psi
_{2},\Psi_{1}\right\rangle \right)  \left\vert g\right\rangle . \label{State3}%
\end{align}
Finally, by measuring $\left\vert m\right\rangle $\ and $\left\vert
g\right\rangle $ (see the method in the following example), we make the
photonic sector of the total state collapse to the target symmetric entangled
states $\left\vert \Psi_{1},\Psi_{2}\right\rangle \pm\left\vert \Psi_{2}%
,\Psi_{1}\right\rangle $. For clarity the complete procedure of the above is
summarized in Fig. 1.

A candidate for the ancilla system should satisfy two requirements. First, the
quantum system should have two long-lived and well separated states between
which a $\sigma^{x}$\ rotation can be performed. The second requirement is
specified by the swapping stage---the system should have an appropriate energy
level structure for the formation of the TWM or FWM interaction loop where two
of the transitions have to be strongly coupled to the input fields. These
conditions can be satisfied by certain trapped natural atoms or ions, single
color centers, quantum dots or superconducting qubits based on the Josephson
junctions, which have multi-level structures and can also be strongly coupled
to the suitable field modes.

Different from the idea of inducing the conditional interaction of matter wave
(ion or atom) state superposition with one optical mode \cite{catTheory} for
creating the cat states \cite{cat1, cat2,cat3,gerry}, our setup allows to
realize a conditional coupling directly between two photonic modes for their
swapping. This is necessary for constructing a SECS $|\alpha,\beta\rangle
\pm|\beta,\alpha\rangle$ with $\alpha\gg\beta$ (for making a cat state of
large size) or a N00N state. Our method aims to generate all such states in a
unified way.

\section{Example of realization}

Here we provide an example to implement the protocol with the single ion of
calcium $Ca^{+}$ trapped in ion trap and embedded in optical resonator as the
ancilla system. The energy level structure of the calcium ion $Ca^{+}$\ is
illustrated in Fig. 2(a). Trapped ions are well studied systems for quantum
information processing \cite{ion}. The construction of multi-partite entangled
states of trapped ions themselves has been proposed in \cite{lamata}. Our
proposed setup for entangling cavity fields via a trapped ion is similar to
that of the recent experiments reported in \cite{blatt, Stute2012}.%
\begin{figure}
[ptb]
\begin{center}
\includegraphics[
height=2.1465in,
width=3.5008in
]%
{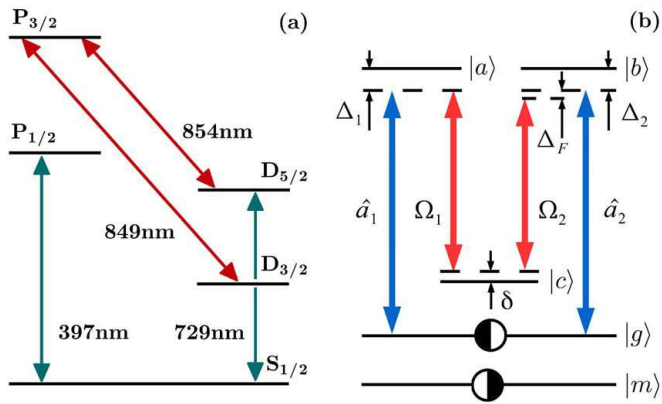}%
\caption{(color online). (a) $Ca^{+}$ energy level scheme. (b) Level scheme
for realizing the conditional BS coupling between two photonic modes
$\widehat{a}_{1}$ and $\widehat{a}_{2}$ (blue arrows). The red arrows
represent the classical coupling fields. The white and black circles on levels
$\left\vert g\right\rangle $ and $\left\vert m\right\rangle $ indicate that
they are in a superposition and only one of them is involved into the
parametric loop. The energy levels of this general scheme correspond to those
of $Ca^{+}$ for our example as follows: $\left\vert m\right\rangle
\rightarrow$ $4S_{1/2}\left(  F=4,~m_{J}=0\right)  $; $\left\vert
g\right\rangle \rightarrow$ $3D_{5/2}\left(  F=6,~m_{J}=0\right)  $;
$\left\vert a\right\rangle \rightarrow$ $4P_{3/2}\left(  F=5,~m_{J}=1\right)
$; $\left\vert b\right\rangle \rightarrow$ $4P_{3/2}\left(  F=5,~m_{J}%
=-1\right)  $ and $\left\vert c\right\rangle \rightarrow$ $3D_{3/2}\left(
F=5,~m_{J}=0\right)  $.}%
\end{center}
\end{figure}
The two ground states $\left\vert m\right\rangle $ and $\left\vert
g\right\rangle $\ of the ancilla ion we use are $4S_{1/2}\left(
F=4,~m_{J}=0\right)  $\ and $3D_{5/2}\left(  F=6,~m_{J}=0\right)  $,
respectively. These particular levels are chosen as the ground states for two
reasons. First, both of the states are long-lived (up to the order of $1$ s).
Second, due to the selection rule and large energy difference between them,
one of the ground states is excluded from our parametric FWM loop to swap two
photonic modes so that the conditional BS coupling can be realized.

The photonic modes we will swap are prepared as the steady-state fields of the
cavity. The process building up the steady cavity field can be described by
the Hamiltonian ($\hbar\equiv1$)
\begin{align}
H_{c}  &  =\sum_{j=1}^{2}\big\{iE_{j}(\hat{a}_{j}^{\dagger}e^{-i\Delta_{0j}%
t}-\hat{a}_{j}e^{i\Delta_{0j}t})\nonumber\\
&  +i\sqrt{\kappa}\big(\hat{a}_{j}^{\dagger}\hat{\xi}_{c}(t)-\hat{a}_{j}%
\hat{\xi}_{c}^{\dagger}(t)\big)\big\} \label{cavity}%
\end{align}
in the interaction picture with respect to the cavity Hamiltonian $\sum
_{j}\omega_{cj}\hat{a}_{j}^{\dagger}\hat{a}_{j}$. The first term in the above
describes the continuous wave (CW) drives of the frequency $\omega_{0j}$, and
with the intensity $E_{j}$ and the detuning $\Delta_{0j}=\omega_{cj}%
-\omega_{0j}$ from the cavity frequency $\omega_{cj}$. The second term about
the coupling between the cavity modes and the cavity noise operator $\hat{\xi
}_{c}$ gives rise to the damping of the cavity at the rate $\kappa$
\cite{noise, h-12-2}. The steady cavity fields in the coherent state with
$|\langle\hat{a}_{j}\rangle|=E_{j}/(0.5\kappa+i\Delta_{0j})$ will be created
by driving the cavity for a while. The cavity fields in Fock state can be
established by the technique in \cite{LawEberly}. The two prepared
intra-cavity modes in the state $\left\vert \Psi_{1}\right\rangle _{\sigma
^{+}}$ and $\left\vert \Psi_{2}\right\rangle _{\sigma^{-}}$ of the different
polarization will be coupled to the transition $3D_{5/2}\left(  F=6\right)
\leftrightarrow P_{3/2}\left(  F=5\right)  $ at $854$ nm of the trapped ion.

In case when the intra-cavity modes are in the coherent states the SECSs will
be generated and for the Fock and vacuum inputs the N00N states will be
obtained as the result of the conditional swapping. As follows from the
interaction configuration presented in Fig.2(b) only the ground state
$\left\vert g\right\rangle $ ($3D_{5/2}\left(  F=6,~m_{J}=0\right)  $) becomes
coupled to the optical modes but there is no coupling to the optical modes for
the state $\left\vert m\right\rangle $ ($4S_{1/2}\left(  F=4,~m_{J}=0\right)
$). In order to perform a $\sigma^{x}$ rotation between $\left\vert
m\right\rangle $ and $\left\vert g\right\rangle $ and bring the ion into the
superposition state in Eq. (\ref{State1}), a resonant $\pi/2$ laser pulse at
$729$ nm with\ $\pi$ - polarization is applied to the quadrupole transition
$4S_{1/2}\leftrightarrow3D_{5/2}$. During the swapping stage there are also
two classical pumping pulses with the orthogonal circular polarizations
applied to the transitions $3D_{3/2}\left(  F=5,m_{J}=0\right)
\leftrightarrow4P_{3/2}\left(  F=5,m_{J}=\pm1\right)  $ at $849$ nm, while the
cavity modes $\sigma^{+}$\ and $\sigma^{-}$ are coupled to the transitions
$3D_{5/2}\left(  F=6,m_{J}=0\right)  \leftrightarrow4P_{3/2}\left(
F=5,m_{J}=\pm1\right)  $ in the parametric FWM loop.

To realize the parametric BS coupling, all real transitions in the loop should
be suppressed and ideally the ion should stay in its ground state during the
swapping process. Therefore all fields should be highly detuned from the
resonance and satisfy certain conditions (see the discussion below). By
controlling the duration of the classical pulses we can control the precise
parametric interaction time for obtaining the state in (\ref{State2}). After
another $\pi/2$ laser pulse at $729$ nm with\ $\pi$ - polarization there will
be the state in (\ref{State3}). The detection of the ground states for the
final projection onto the target states is implemented by exciting the
transition $4S_{1/2}\rightarrow4P_{1/2}$ at $397$ nm; see the similar
technique in \cite{detect, detect1}. The presence of the fluorescence
collapses the ion wave function onto $\left\vert m\right\rangle $ and the
absence of the fluorescence indicates the state $\left\vert g\right\rangle $.

\section{mechanism for induced Beam-splitter coupling}

The dispersive FWM process for realizing the conditional swapping in our
protocol can be implemented in any system with the level scheme in Fig. 2(b).
The Hamiltonian for the process shown in Fig. 2(b) takes the form
($\hbar\equiv1$)
\begin{align}
H  &  =-\Delta_{1}\sigma_{aa}-\Delta_{2}\sigma_{bb}-\delta\sigma_{cc}%
+g_{1}\widehat{a}_{1}^{\dagger}\sigma_{ga}\nonumber\\
&  +g_{2}\widehat{a}_{2}^{\dagger}\sigma_{gb}+\Omega_{1}\sigma_{ca}+\Omega
_{2}e^{i\Delta_{F}t}\sigma_{cb}+h.c. \label{coupling}%
\end{align}
in a rotated frame. Here $\sigma_{ij}=\left\vert i\right\rangle \left\langle
j\right\vert $ is the atomic spin flip operator; $g_{1\left(  2\right)  }$ is
the coupling constant; $\Delta_{1(2)}=\omega_{a1(2)}-\omega_{ga(b)}$ is the
one-photon detuning, $\delta=\omega_{a1}-\omega_{1}-\omega_{cg}$ is the
two-photon detuning, and $\Delta_{F}=\omega_{a1}-\omega_{1}-\omega_{a2}%
+\omega_{2}$ is the four-photon detuning, with $\omega_{1(2)}$ being the
frequency of the classical pumping pulse with the Rabi frequency
$\Omega_{1(2)}$, $\omega_{a1(a2)}$ being the frequency of the input pulse. The
modes $\hat{a}_{i}$ are the steady field intra-cavity modes for the example
described in the last section. Given the possibility to prepare the many-body
superposition $(|m_{1},\cdots,m_{n}\rangle-i|g_{1},\cdots,g_{n}\rangle
)/\sqrt{2}$ of an ensemble of the atoms with a similar level scheme to
Fig.2(b), the dispersive FWM process can also be performed in the ensemble.
Then $\hat{a}_{i}$ are just the representative modes of the narrow band input pulses.

The Hamiltonian (\ref{coupling}) only presents the coherent part of the
interaction process without dissipations. In Sec. VI we will give the detailed
discussion on the decoherence effects arising due to the decay of the energy
levels of the ancilla system and the loss of the cavity.

For the process in Fig.2(b) the Schr\"{o}dinger equation for each energy level
component of the state $|\Psi(t)\rangle$ reads:
\begin{subequations}
\begin{equation}
i\frac{d}{dt}\left\langle g|\Psi\left(  t\right)  \right\rangle =g_{1}%
\widehat{a}_{1}^{\dagger}\left\langle a|\Psi\left(  t\right)  \right\rangle
+g_{2}\widehat{a}_{2}^{\dagger}\left\langle b|\Psi\left(  t\right)
\right\rangle \label{g_amp}%
\end{equation}%
\begin{align}
i\frac{d}{dt}\left\langle b|\Psi\left(  t\right)  \right\rangle  &
=-\Delta_{2}\left\langle b|\Psi\left(  t\right)  \right\rangle \nonumber\\
+  &  \Omega_{2}e^{-i\Delta_{F}t}\left\langle c|\Psi\left(  t\right)
\right\rangle +g_{2}\widehat{a}_{2}\left\langle g|\Psi\left(  t\right)
\right\rangle \label{b_amp1}%
\end{align}%
\begin{align}
i\frac{d}{dt}\left\langle a|\Psi\left(  t\right)  \right\rangle  &
=-\Delta_{1}\left\langle a|\Psi\left(  t\right)  \right\rangle \nonumber\\
&  +\Omega_{1}\left\langle c|\Psi\left(  t\right)  \right\rangle
+g_{1}\widehat{a}_{1}\left\langle g|\Psi\left(  t\right)  \right\rangle
\label{a_amp}%
\end{align}%
\begin{align}
i\frac{d}{dt}\left\langle c|\Psi\left(  t\right)  \right\rangle  &
=-\delta\left\langle c|\Psi\left(  t\right)  \right\rangle +\Omega
_{1}\left\langle a|\Psi\left(  t\right)  \right\rangle \nonumber\\
&  +\Omega_{2}e^{i\Delta_{F}t}\left\langle b|\Psi\left(  t\right)
\right\rangle . \label{e_amp}%
\end{align}
The effective BS Hamiltonian for the similar dispersive FWM schemes can be
derived by the time-independent perturbation method
\cite{SharypovFWMSuperconductors}. Here we apply the more general method of
the adiabatic elimination \cite{GardinerThesis} to show the realization of the
effective BS coupling. It is important to mention that the one- and two-photon
detunings should be high enough to prevent any real transition of the system
from its ground state. It is therefore possible to see the effective dynamics
of the photonic modes while the system is staying in the ground state
$|g\rangle$.

First, assuming that initially the system is prepared in its ground state
$\left\vert g\right\rangle $, i.e. $\left\langle b|\Psi\left(  t_{0}\right)
\right\rangle =\left\langle a|\Psi\left(  t_{0}\right)  \right\rangle
=\left\langle c|\Psi\left(  t_{0}\right)  \right\rangle =0$, we eliminate the
transitions from the state $\left\vert g\right\rangle $ to $\left\vert
a\right\rangle $ and $\left\vert b\right\rangle $. Under this assumption we
integrate Eqs. (\ref{g_amp}) - (\ref{a_amp}) and then substitute the formal
solution of $\left\langle g|\Psi\left(  \tau\right)  \right\rangle $ into
those of $\left\langle b|\Psi\left(  t\right)  \right\rangle $ and
$\left\langle a|\Psi\left(  t\right)  \right\rangle $ to obtain the relations%
\end{subequations}
\begin{subequations}
\begin{equation}
\left\langle b|\Psi\left(  t\right)  \right\rangle =\frac{\Omega
_{2}e^{-i\Delta_{F}t}}{\Delta_{2}}\left\langle c|\Psi\left(  t\right)
\right\rangle +\frac{g_{2}\widehat{a}_{2}}{\Delta_{2}}\left\langle
g|\Psi\left(  t\right)  \right\rangle \label{b5}%
\end{equation}%
\begin{equation}
\left\langle a|\Psi\left(  t\right)  \right\rangle =\frac{\Omega_{1}}%
{\Delta_{1}}\left\langle c|\Psi\left(  t\right)  \right\rangle +\frac
{g_{1}\widehat{a}_{1}}{\Delta_{1}}\left\langle g|\Psi\left(  t\right)
\right\rangle , \label{aa5}%
\end{equation}
where we are concerned with the regime of $\left\vert g_{i}\sqrt{n_{i}}%
/\Delta_{i}\right\vert \ll1$ and keep only the first order of this small term,
and $n_{i}$ is the average photon number of the $i$-th input mode.

Next, in order to obtain the decoupled dynamics of the effective two-level
system of $\left\vert g\right\rangle $ and $\left\vert c\right\rangle $, we
substitute Eqs.(\ref{b5}) and (\ref{aa5}) into Eqs. (\ref{g_amp}) and
(\ref{e_amp}) and obtain%
\end{subequations}
\begin{subequations}
\begin{equation}
\frac{d}{dt}e^{i\alpha_{g}t}\left\langle g|\Psi\left(  t\right)  \right\rangle
=-ie^{i\alpha_{g}t}\beta^{\dagger}\left(  t\right)  \left\langle c|\Psi\left(
t\right)  \right\rangle \label{g_TLS}%
\end{equation}%
\begin{equation}
\frac{d}{dt}e^{i\alpha_{c}t}\left\langle c|\Psi\left(  t\right)  \right\rangle
=-ie^{i\alpha_{c}t}\beta\left(  t\right)  \left\langle g|\Psi\left(  t\right)
\right\rangle , \label{e_TLS}%
\end{equation}
where we have introduced the functions $\alpha_{c}=-\delta+\Omega_{1}%
^{2}/\Delta_{1}+\Omega_{2}^{2}/\Delta_{2}$, $\alpha_{g}=g_{1}^{2}\widehat
{a}_{1}^{\dagger}\widehat{a}_{1}/\Delta_{1}+g_{2}^{2}\widehat{a}_{2}^{\dagger
}\widehat{a}_{2}/\Delta_{2}$, and $\beta=\Omega_{1}g_{1}\widehat{a}_{1}%
/\Delta_{1}+\Omega_{2}g_{2}\widehat{a}_{2}e^{i\Delta_{F}t}/\Delta_{2}$. In the
dynamics of this effective two-level system the parameter $\beta$ plays the
role of the effective coupling constant and the parameter%
\end{subequations}
\begin{equation}
\Delta_{eff}=\alpha_{g}-\alpha_{c}\approx\delta-\Omega_{1}^{2}/\Delta
_{1}-\Omega_{2}^{2}/\Delta_{2} \label{DetuningTLS}%
\end{equation}
corresponds to the effective detuning. We have considered the regime
satisfying $\left\vert g_{i}^{2}n_{i}/\Delta_{i}\right\vert \ll\left\vert
\delta-\Omega_{1}^{2}/\Delta_{1}-\Omega_{2}^{2}/\Delta_{2}\right\vert $ in
(\ref{DetuningTLS}). The dynamics of the states $\left\vert g\right\rangle $
and $\left\vert c\right\rangle $ will be decoupled further. Integrating
Eq.(\ref{e_TLS}) we get the relation%
\begin{equation}
\left\langle c|\Psi\left(  t\right)  \right\rangle =\frac{1}{\Delta_{eff}%
}\left[  \frac{\Omega_{1}g_{1}\widehat{a}_{1}}{\Delta_{1}}+\frac{\Omega
_{2}g_{2}\widehat{a}_{2}}{\Delta_{2}}e^{i\Delta_{F}t}\right]  \left\langle
g|\Psi\left(  t\right)  \right\rangle , \label{e_eliminated}%
\end{equation}
where we keep only the first order of the parameter $\left\vert g_{i}%
\sqrt{n_{i}}\Omega_{i}/\left(  \Delta_{i}\Delta_{eff}\right)  \right\vert
\ll1$.

Finally, substituting Eq. (\ref{e_eliminated}) into (\ref{g_TLS}), we obtain
the decoupled evolution of the state $\left\vert g\right\rangle $%
\begin{equation}
i\frac{d}{dt}\left\langle g|\Psi\left(  t\right)  \right\rangle =H_{eff}%
\left\langle g|\Psi\left(  t\right)  \right\rangle , \label{g_decoupled}%
\end{equation}
where
\begin{equation}
H_{eff}(t)=\eta_{1}\widehat{a}_{1}^{\dagger}\widehat{a}_{1}+\eta_{2}%
\widehat{a}_{2}^{\dagger}\widehat{a}_{2}+\chi_{0}\left(  \widehat{a}_{1}%
^{\dag}\widehat{a}_{2}e^{i\Delta_{F}t}+h.c.\right)  \label{H_eff_disp_FWM}%
\end{equation}
is the effective Hamiltonian for the photonic modes, with $\eta_{i}%
=\frac{g_{i}^{2}}{\Delta_{i}}+\frac{g_{i}^{2}\Omega_{i}^{2}}{\Delta_{i}%
^{2}\Delta_{eff}}$, $\chi_{0}=\frac{\Omega_{1}\Omega_{2}g_{1}g_{2}}{\Delta
_{1}\Delta_{2}\Delta_{eff}}$.

The conditions $\left\vert g_{i}\sqrt{n_{i}}/\Delta_{i}\right\vert \ll1$,
$\left\vert g_{i}\sqrt{n_{i}}\Omega_{i}/\left(  \Delta_{i}\Delta_{eff}\right)
\right\vert \ll1$ leading to the above effective dynamics prevent the one- and
two-photon transitions out of a ground level and can be realized by adjusting
the system parameters. For our example using $Ca^{+}$ with $g_{1(2)}\sim10$
MHz, it is possible to set the Rabi frequencies $\Omega_{1(2)}\sim1$ GHz, the
one-photon detunings $\Delta_{1(2)}\sim1$ GHz, and the two-photon detuning
$\delta\sim1$ GHz, given the average photon numbers up to $n_{i}=100$. The
symbol \textquotedblleft$\sim$" means the order of the values here. The sizes
$\sqrt{n_{i}}$ of the states to be entangled can be made larger simply by
increasing the detunings.

\section{performance of Swapping operation}

The unitary evolution operator of the time-dependent effective Hamiltonian
$H_{eff}$ in (\ref{H_eff_disp_FWM}) can be decomposed as
\begin{align}
&  \mbox{T}\exp\left\{  -i\int_{0}^{t}d\tau H_{eff}(\tau)\right\}
=\exp\left\{  -i\eta_{1}\widehat{a}_{1}^{\dag}\widehat{a}_{1}t-i\eta
_{2}\widehat{a}_{2}^{\dag}\widehat{a}_{2}t\right\} \nonumber\\
&  \times\mbox{T}\exp\left\{  -i\chi_{0}\int_{0}^{t}e^{-i\delta_{F}\tau}%
d\tau~\widehat{a}_{1}\widehat{a}_{2}^{\dag}+h.c.\right\}  , \label{decomp}%
\end{align}
where $\mbox{T}$ stands for a time-ordered operation and $\delta_{F}%
=\Delta_{F}+\eta_{1}-\eta_{2}$. The general form of such decomposition is
given in \cite{h-12-2}. The first of the decomposed operators in
Eq.(\ref{decomp}) is a phase shift operation and the second is a BS operation.
For example, by tuning the system parameters so that the conditions
$\delta_{F}=0$ and $\Delta_{eff}=-2\Delta_{i}$ (assuming $g_{1}=g_{2}$ and
$\frac{\Omega_{1}}{\Delta_{1}}=\frac{\Omega_{2}}{\Delta_{2}}=1$) are
satisfied, their combined action implements an ideal swapping $\widehat{a}%
_{1}\leftrightarrow\widehat{a}_{2}$ after the time $t_{s}$ accumulating
$|\chi_{0}t_{s}|=0.5\pi$. Given the data following Eq.(\ref{H_eff_disp_FWM}) a
pair of input states could be entangled within a few microseconds.%

\begin{figure}
[ptb]
\begin{center}
\includegraphics[
height=1.4927in,
width=3.416in
]%
{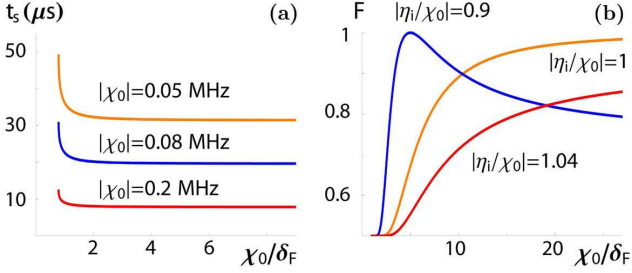}%
\caption{(color online). (a) Relation between the swapping time and the ratio
$\chi_{0}/\delta_{F}$. (b) Fidelity of the generated states with the target
output $|\alpha\rangle|\beta\rangle\pm|\beta\rangle|\alpha\rangle$ with
$\alpha=6\sqrt{2}$ and $\beta=\sqrt{2}$, which can be converted to the cat
states $|\gamma\rangle\pm|-\gamma\rangle$ of $\gamma=5$. A unit fidelity can
be reached in the regime $|\eta_{i}/\chi_{0}|<1$ at a not so large ratio
$\chi_{0}/\delta_{F}$.}%
\end{center}
\end{figure}
In a general situation the swapping time $t_{s}$ is determined by the relation
$2\chi_{0}\sin(0.5\delta_{F}t_{s})/\delta_{F}=0.5\pi$, implying a quickly
stabilized swapping time with increasing ratio $\chi_{0}/\delta_{F}$; see Fig.
3(a). Meanwhile the output state will be $|\alpha,\beta\rangle\pm|\beta
e^{i\varphi_{1}},\alpha e^{i\varphi_{2}}\rangle$, where $\varphi_{i}=\eta
_{i}t_{s}-0.5\pi+0.5\delta_{F}t_{s}$, if the inputs are two coherent states
$|\alpha\rangle$ and $|\beta\rangle$. The fidelity of the output state is
determined by the two ratios $|\chi_{0}/\delta_{F}|$ and $|\eta_{i}/\chi_{0}|$
(given $\eta_{1}=\eta_{2}$). As it is shown in Fig. 3(b), a high-quality
output state will be realized with the proper ratios that can be achieved by
adjusting the system parameters.

The frequency dependent parameters in the effective Hamiltonian
(\ref{H_eff_disp_FWM}) might also decrease the BS fidelity when one considers
the multi-frequency pulses. In fact, such difference is negligible to the
narrow-band input pulses and can be avoided by using the interaction scheme
for the BS transformation \cite{lin-08} where the effective Hamiltonian is
insensitive to the frequencies of the input modes.

\section{Decoherence effects}

For the ancilla system applied in the parametric loop, the decoherence effect
is from two sources. One is due to the broadening of the transitions, so the
fields detuning $\Delta_{i}$, $\Delta_{eff}$ should be much larger than the
bandwidth of the corresponding transitions $\Gamma_{gi}$ ($i=a,b$ and $c$),
i.e. $\left\vert \Delta_{i}\right\vert \gg\Gamma_{ga,gb}$ and $\left\vert
\Delta_{eff}\right\vert \gg\Gamma_{gc}$. The other more significant one comes
from the possible population of the other energy levels than the ground state
$|g\rangle$\ during the swapping process. The radiative decay in the system
occurs only after the excitation of the system from the ground state. As we
apply a dispersive interaction, the probability for the one-photon excitation
scales as $P_{a,b}=g_{1,2}^{2}n_{1,2}\Delta_{1,2}^{-2}$ (see Eqs. (\ref{b5})
and (\ref{aa5})), and the probability of the two-photon excitation scales as
$P_{c}=|\Omega_{1}g_{1}\sqrt{n_{1}}/(\Delta_{1}\Delta_{eff})+\Omega_{2}%
g_{2}\sqrt{n_{2}}e^{i\Delta_{F}t}/(\Delta_{2}\Delta_{eff})|^{2}$ (see Eq.
(\ref{e_eliminated})). Therefore the effective decays in the system should be
described by the product of the decay rate of the particular state and the
probability of the electron excitation at this level. In order to have the
negligible contribution from the decay processes, the swapping time should be
much shorter than the effective radiative lifetime $t_{s}\ll1/(\gamma_{i}%
P_{i})$. For example, if one takes the swapping time for the matched
four-photon detuning giving $\delta_{F}=0$, there should be the condition
$\gamma_{a,b}\ll\frac{2}{\pi}\left\vert \frac{\Omega_{1}\Omega_{2}}%
{\Delta_{eff}}\frac{\Delta_{1,2}}{\Delta_{2,1}}\frac{g_{2,1}}{g_{1,2}%
}\right\vert $ for neglecting the decay of the intermediate states $\left\vert
a\right\rangle $ ($\left\vert b\right\rangle $) and the condition $\gamma
_{c}\ll\frac{2}{\pi}|\sqrt{n_{1}m}+\sqrt{n_{2}e^{i\Delta_{F}t}m^{-1}}%
|^{-2}\left\vert \Delta_{eff}\right\vert $, where $m=\frac{\Omega_{1}%
g_{1}\Delta_{2}}{\Omega_{2}g_{2}\Delta_{1}}$, for neglecting the decay of the
state $\left\vert c\right\rangle $. For the example using $Ca^{+}$ with
$\gamma_{a,b}\sim10$ MHz and $\gamma_{c}\sim10$ Hz the above conditions can be
easily established. These conditions guarantee the outputs of the swapping
process to be pure states.

Another type of decoherence arises from the loss of steady cavity fields in
any cavity based implementation. It modifies the unitary evolution in
(\ref{decomp}) to a non-unitary one governed by the mater equation
\[
\dot{\rho}=-i[H_{eff},\rho]+\sum_{i}\kappa\big(\hat{a}_{i}\rho\hat{a}%
_{i}^{\dagger}-\frac{1}{2}(\rho\hat{a}_{i}^{\dagger}\hat{a}_{i}+\hat{a}%
_{i}^{\dagger}\hat{a}_{i}\rho)\big),
\]
where $H_{eff}$ is given in (\ref{H_eff_disp_FWM}). Given the input as a
product of two coherent states $|\alpha\rangle|\beta\rangle$, for example, the
solution to the master equation for a short evolution time and under the
condition of reaching the unit fidelity in Fig. 3(b) can be found by following
a similar procedure in \cite{h-09} as
\begin{align}
\rho(t)  &  =\frac{1}{N_{\alpha,\beta}}\big(|\alpha e^{-\kappa t/2},\beta
e^{-\kappa t/2}\rangle\langle\alpha e^{-\kappa t/2},\beta e^{-\kappa
t/2}|\nonumber\\
&  +|\beta e^{-\kappa t/2},\alpha e^{-\kappa t/2}\rangle\langle\beta
e^{-\kappa t/2},\alpha e^{-\kappa t/2}|\nonumber\\
&  +C|\alpha e^{-\kappa t/2},\beta e^{-\kappa t/2}\rangle\langle\beta
e^{-\kappa t/2},\alpha e^{-\kappa t/2}|\nonumber\\
&  +C|\beta e^{-\kappa t/2},\alpha e^{-\kappa t/2}\rangle\langle\alpha
e^{-\kappa t/2},\beta e^{-\kappa t/2}|,
\end{align}
where $\sqrt{N_{\alpha,\beta}}$ is the normalization factor for the SECS
$|\alpha\rangle|\beta\rangle+|\beta\rangle|\alpha\rangle$, and $C=\exp
\{-2|\alpha-\beta|^{2}(1-e^{-\kappa t})\}$. Such decoherence could be
negligible to a cavity of high finesse. Using a cavity with the damping time
$0.13$ seconds as in \cite{cat3}, for example, the fidelity (with a pure SECS)
of the generated state for the example in Fig. 3(b) can be preserved over
$0.98$ up to the conditional swapping time $0.1$ milliseconds.

\section{conclusion}

With an example, we have illustrated how to entangle two arbitrary states
$|\Psi_{1}\rangle$ and $|\Psi_{2}\rangle$ to the symmetric form $\left\vert
\Psi_{1},\Psi_{2}\right\rangle \pm\left\vert \Psi_{2},\Psi_{1}\right\rangle $
with an induced conditional BS type coupling that avoids the physical
limitation on nonlinear couplings. In contrast to all previous schemes, the
entanglement strategy is independent of the specific input states, e.g. SECSs
and N00N states can be generated in the same way. The FWM process for
realizing the effective BS coupling is within the current experimental
technology. This approach can help to achieve the goals of entangling light
fields with flexibility and creating cat states of large size.

\acknowledgements
A. V. S. was supported by RFBR 12-02-31621, and B. H. acknowledges the support
by AITF.

\end{document}